\documentclass{ws-procs9x6}

\newcommand{\tc}{T_c}
\newcommand{\jpsi}{J / \psi}
\newcommand{\ec}{\eta_c}

\newcommand{\gt}{G(\tau, \vec{p}, T)}
\newcommand{\grecon}{G_{{\rm recon}, T^*}(\tau, \vec{p}, T)}
\newcommand{\sgw}{\sigma(\omega, \vec{p}, T)}

\begin{document}

\title{Charmonia at finite momenta in a deconfined plasma
\footnote{\uppercase{T}his research is supported by \uppercase{BMBF} under 
grant no. \uppercase{06BI102}, and by \uppercase{DOE} under contract 
\uppercase{DE-AC02-98CH10886}.}}

\author{S. Datta, F. Karsch and S. Wissel} 

\address{Fakult\"at f\"ur Physik, Universit\"at Bielefeld, 
D-33615 Bielefeld, Germany. \\ 
E-mail: saumen,karsch,wissel@physik.uni-bielefeld.de}

\author{P. Petreczky \footnote{\uppercase{G}oldhaber and 
\uppercase{RIKEN} fellow}}

\address{Brookhaven National Laboratory, Upton, NY 11973, USA.
\\ E-mail: petreczk@quark.phy.bnl.gov}  

\author{I. Wetzorke}

\address{NIC/DESY Zeuthen, Platanenallee 6, D-15738 Zeuthen, Germany.
\\ E-mail: Ines.Wetzorke@desy.de}

\maketitle

\abstracts{Lattice studies of charmonium systems have indicated that in
a deconfined gluonic plasma ground state charmonia survive as bound states 
upto temperatures $\sim 2 \tc$. After surveying the methodologies used 
in reaching these results, we examine the behavior of these systems when the 
bound state is in motion with respect to the heatbath frame. We find that
the finite momenta charmonia show medium modifications when the medium
is deconfined; in particular, a modification of the energy-momentum 
dispersion relation is indicated.}

The study of strongly interacting matter at high temperatures is important
both for understanding QCD and for cosmological applications, and is a very
active field with experimental data available from relativistic heavy
ion collisions.
Charmonium systems have been of great interest to
the relativistic heavy ion community ever since Matsui and
Satz, based on potential model studies, predicted that
charmonium bound states cannot survive in a deconfined
plasma, and suggested $\jpsi$ as a probe of the confinement
status of the plasma \cite{satz}. Later, more detailed potential 
model studies suggested that all charmonium bound states dissolve before
or pretty soon after deconfinement \cite{karsch}. However,
a direct lattice investigation of the hadronic correlators
have revealed that the ground state charmonia ($\ec$ and $\jpsi$) survive the
deconfinement transition in pure gauge theory
\cite{prd,umeda,asakawa}, and may
dissolve only at temperatures $\sim 2 \tc$. 


Both for understanding the mechanism of bound state dissolution in the
plasma and for study of charmonium abundance in experiments, it is important
to also know the fate of these bound states when they not at rest but moving
with respect to the heatbath frame. The lack of covariance in
finite temperature system means the pole of the inverse propagator,
\begin{equation}
\Delta^{-1}(p_0,p,T)=p_0^2+p^2+m^2+{\rm Re} \Pi(p,T)+i {\rm Im} \Pi(p,T)
\label{eq.generic}
\end{equation}
can be a generic function of spatial momentum p, leading possibly to a
modification of the energy-momentum dispersion relation and a momentum
dependent decay width. Such modifications have been discussed in different
models for pions in hadron gas \cite{shuryak}, and for decay width of $\jpsi$ 
in hadron gas \cite{haglin}. Here, we present a preliminary report of a
direct lattice investigation into the fate of the ground state charmonia 
with a finite spatial momentum (in the heatbath frame). In this report, 
we will restrict ourselves to comparatively moderate temperatures 
$\le 1.5 \tc$.

Direct lattice investigations of the fate of the charmonium
states involve analyzing the Matsubara correlators,
\begin{equation}
G_H (\tau, \vec{p}, T) = \langle J_H (\tau, \vec{p}) J_H^\dag
(0, -\vec{p}) \rangle_T
\label{eq.cor}
\end{equation}
where $J_H$ is the suitable mesonic operator, projected on the state with 
spatial momentum $\vec{p}$. For studying the $\ec$ and the $\jpsi$,
we use the point-point operators $\bar{c}
\gamma_5 c$ and $\bar{c} \gamma_\mu c$, 
respectively. For the vector channel, here we
consider the transverse component only.
Through analytical continuation, the
Matsubara correlator can be related to the hadronic spectral function 
by an integral equation:
\begin{equation}
\gt =\int_0^{\infty} d \omega
\sgw \frac{\cosh(\omega(\tau-1/2
T))}{\sinh(\omega/2 T)}.
\label{eq.spect}
\end{equation}

While the inversion of Eq. (\ref{eq.spect}) to get the
spectral function from the correlators measured at
a finite set of points is clearly an ill-defined problem,
some progress has been made in the last few years towards
tackling such problems using Bayesian tools. In the maximum
entropy method \cite{mem} the positivity of the
spectral function and information about the asymptotic
behavior is used to find the most probable spectral
function given data. Analysis of the hadronic correlators 
at low temperatures (below $\tc$) reveal a
complicated structure in the high energy regime of the
spectral functions, dominated by lattice artifacts \cite{prd}. 
We use this structure as part of the asymptotic information in our
analysis at higher temperatures \cite{prd}. 

\begin{figure}[ht]
\centerline{\epsfxsize=2.3in\epsfbox{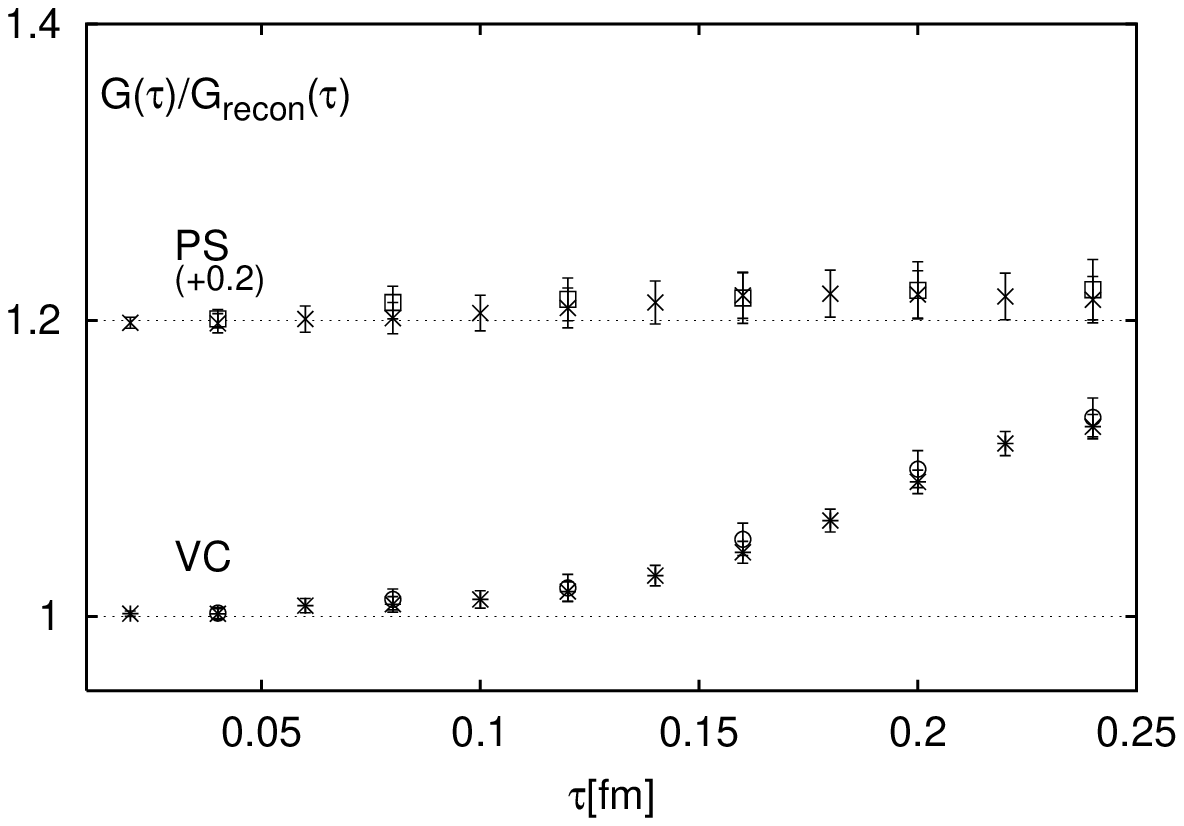}
\epsfxsize=2.3in\epsfbox{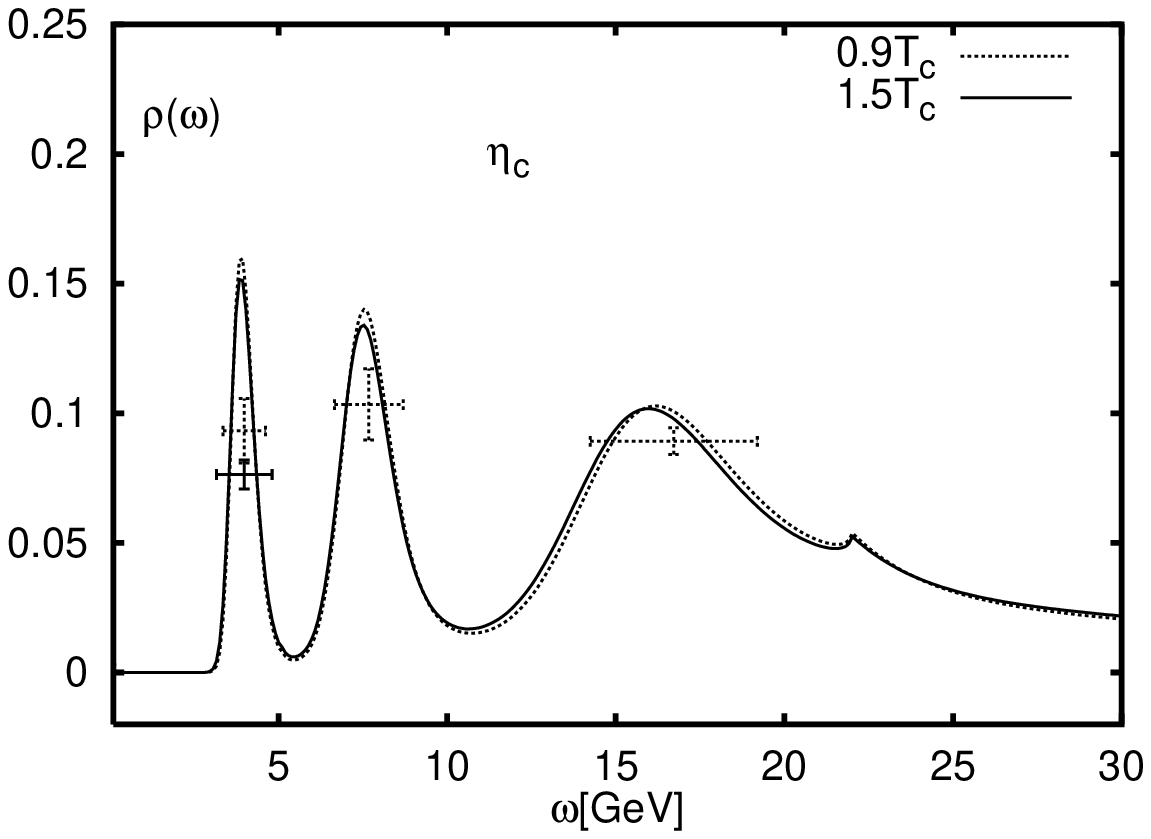}}   
\caption{Comparison of zero momentum correlators at 1.5 $\tc$ for vector
and pseudoscalar charmonia with those reconstructed from
spectral functions at 0.75 $\tc$ (left). The stars and crosses correspond to 
a lattice with lattice spacing of 0.02 fm and the the open symbols correspond
to a lattice with lattice spacing 0.02 fm. (Right) Reconstructed spectral 
functions at 1.5 $\tc$ and 0.9 $\tc$ for the pseudoscalar channel.
Plotted is the rescaled spectral function $\rho(\omega)=
\sigma(\omega) / \omega^2$. The default model included the peak structure 
at high $\omega$, which is dominated by lattice artifacts.}
\label{fig.recon}
\end{figure}

Useful information about possible change of state with
deconfinement can be obtained by comparing the correlators
measured above $\tc$ with the correlators reconstructed
from the spectral function below $\tc$. We define $\grecon$ by substituing
$\sgw$ in Eq. (\ref{eq.spect}) with $\sigma(\omega, \vec{p}, T^*_)$
where $T^*$ is usually taken as the smallest temperature below $\tc$
available to us. This takes into account the
trivial temperature dependence of the kernel and therefore, any deviation
of $\gt$ from $\grecon$ will indicate a temperature
modification of the spectral function \cite{prd}. In Fig.
\ref{fig.recon} the results of such a comparison are
shown for zero momentum correlators. For the pseudoscalar channel, 
Fig. \ref{fig.recon}
shows that the reconstructed correlator explains the
measured correlators completely upto temperatures of 1.5
$\tc$, indicating that there is no significant change for
these states upto this temperature. The spectral function extracted from the 
correlators, shown in the right, confirm this. For $\jpsi$, some small
deviation of the measured correlators is seen from the reconstructed 
correlators, indicating a possible widening \cite{umeda} or mass shift. 
Here we added the three (spatial) components of the vector, and the 
behavior is somewhat different from the four component correlator in Ref. [3]
due to the change in contribution from electric conductivity
below and above $\tc$.

\begin{figure}[ht]
\centerline{\epsfxsize=2.3in\epsfbox{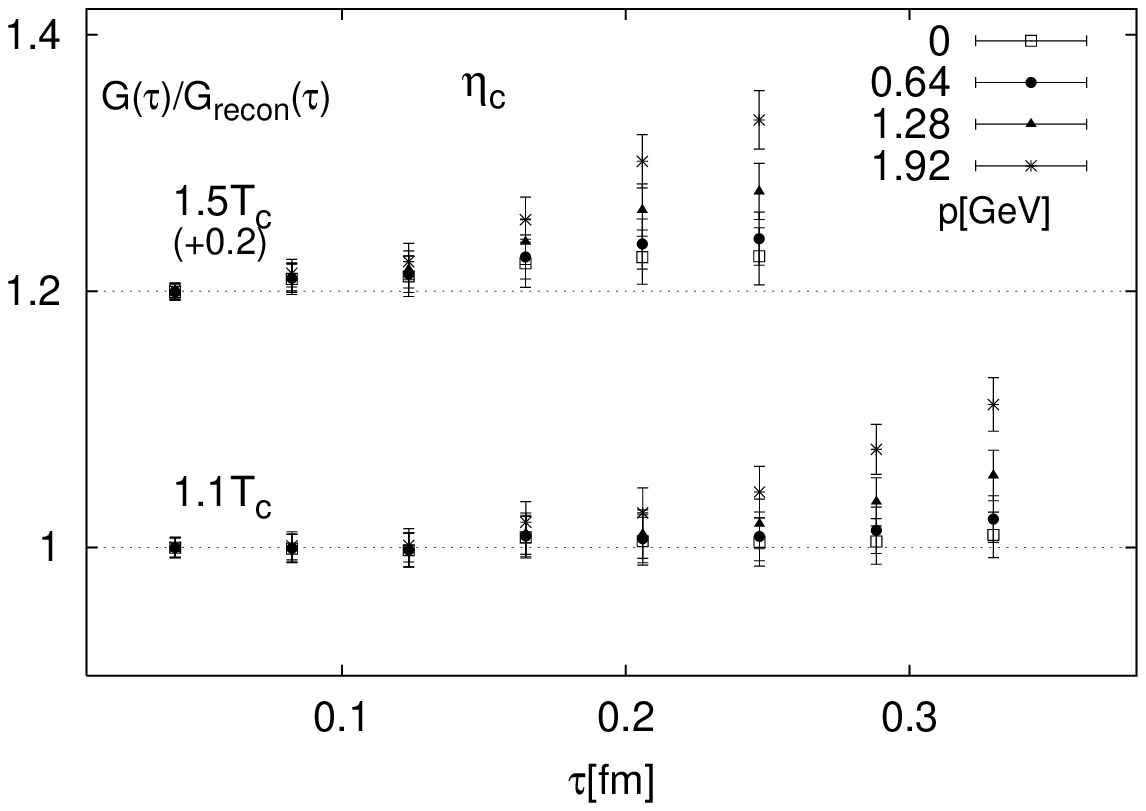}
\epsfxsize=2.3in\epsfbox{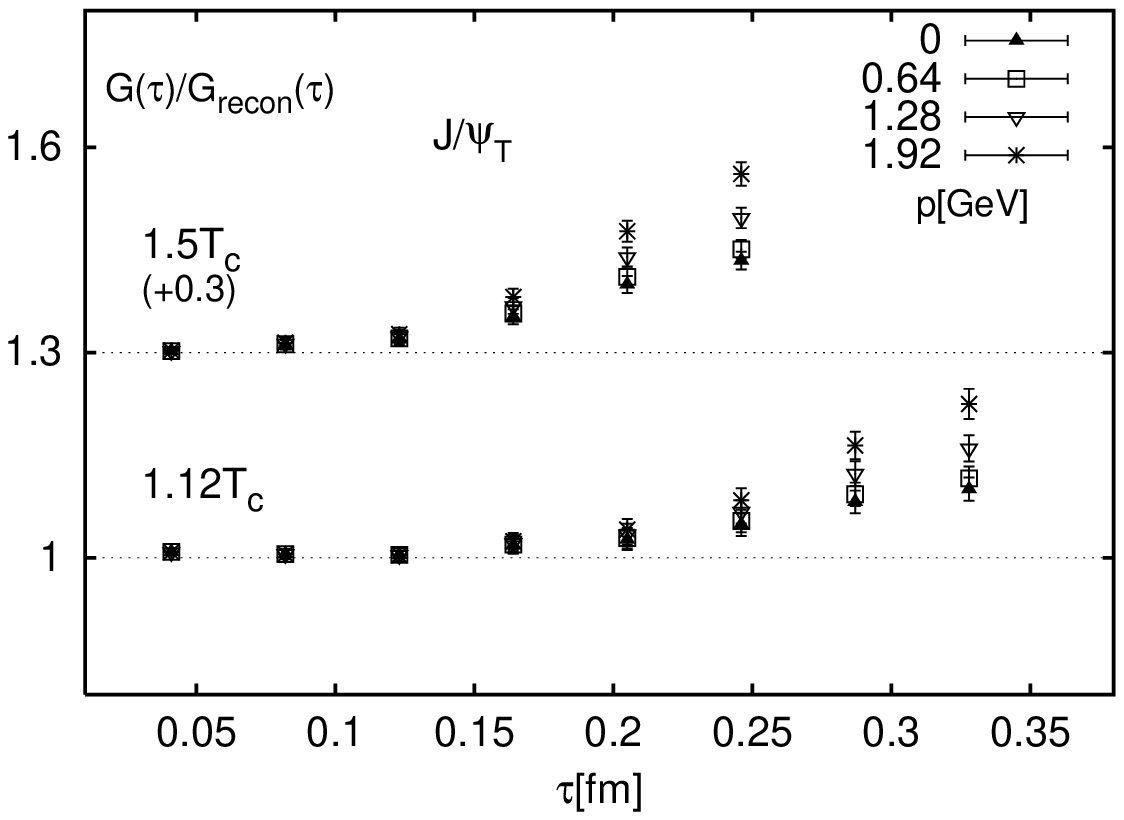}}
\caption{Comparison of correlators at 1.1 $\tc$ and 1.5
$\tc$ for pseudoscalar and (transverse) vector charmonia
with those reconstructed from spectral function at 0.75
$\tc$.}
\label{fig.reconftmom}
\end{figure}

For investigation of the finite momenta charmonia, we use the
same analysis methods as those at zero momentum described
above. Comparison of the finite momentum correlators $\gt$
at 1.1 and 1.5 $\tc$ with the reconstructed correlators from 
$\sigma(\omega, \vec{p}, T)$ are
presented in Fig. \ref{fig.reconftmom}. It can be seen that
with increasing spatial momentum, the correlators above
$\tc$ show larger deviations from the reconstructed
correlators, signalling a medium effect already at 1.1
$\tc$ and increasing with temperature.

\begin{figure}[ht]
\centerline{\epsfxsize=2.3in\epsfbox{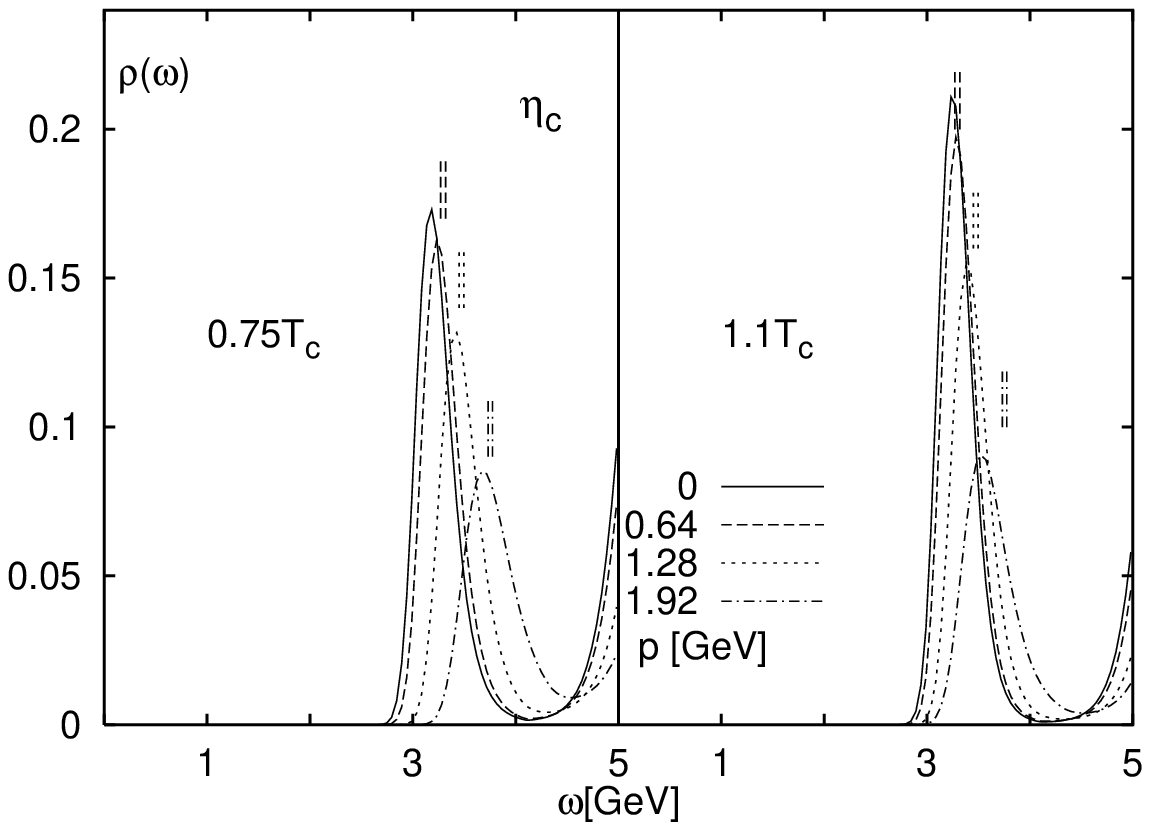}
\epsfxsize=2.3in\epsfbox{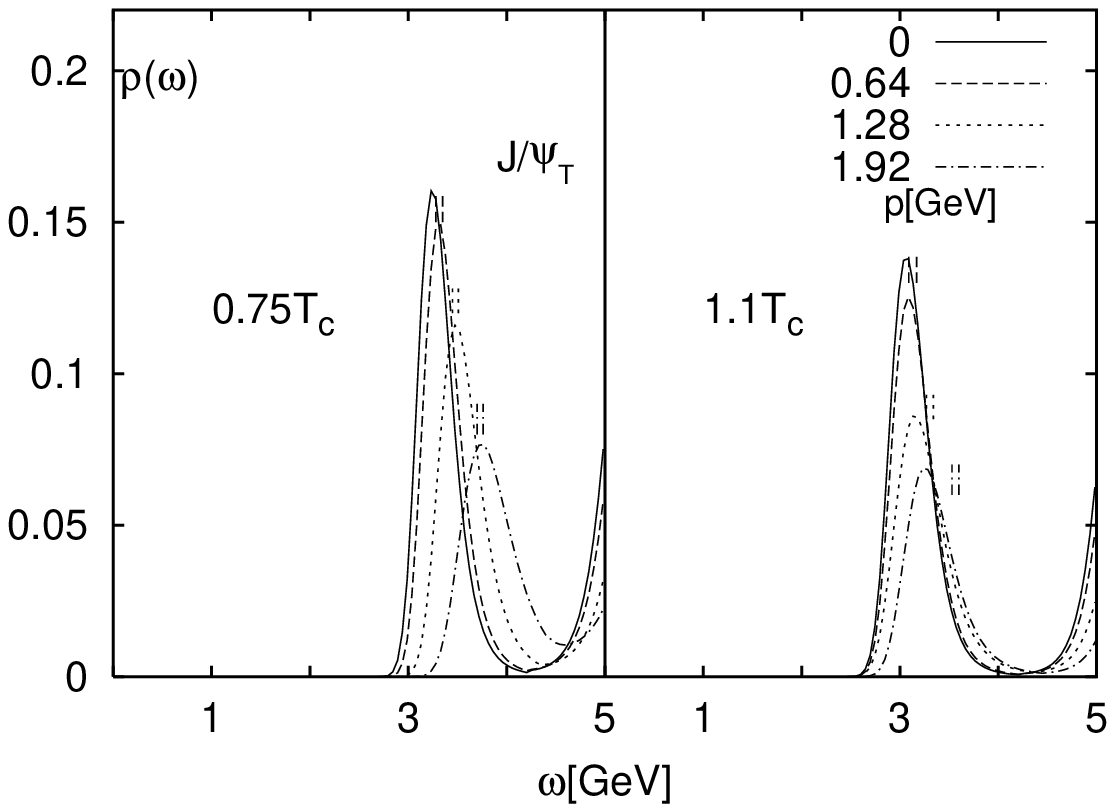}}
\caption{Spectral functions at 0.75 $\tc$ and 1.1 $\tc$ for
different spatial momenta, for lattices with lattice
spacings of 0.04 fm. The small horizontal bars indicate
the expected shift of the peak using relativistic
dispersion relation. The same temporal extent 
has been used at both the temperatures.}
\label{fig.spec664ft}
\end{figure}

In order to get further insight into the nature of the
medium effect, we used maximum entropy analysis for a
direct reconstruction of $\sgw$ above $\tc$. As described
before, for the reconstruction we use the high energy
structure of the spectral function obtained at 0.75 $\tc$
as part of the prior information. The spectral functions at
0.75 $\tc$ and 1.1 $\tc$, using such a default model, are
shown in Fig. \ref{fig.spec664ft}. Here, in order to
avoid any systematics due to the small spatial extent and
limited number of data points at the higher temperature, we
use the same physical extent and same number of data points
both below and above $\tc$, by omitting points at the
center for the lower temperature. 

Figure \ref{fig.spec664ft} shows two interesting
features. First, a strong signal from the ground state shows up in the
spectral function at 1.1 $\tc$ even for momenta $\sim 2
GeV$. Little change in peak strength is seen on
crossing deconfinement, for any of the momenta. Second,
there is an indication of a slower movement of the peak
position above $\tc$, indicating a modification of the
energy-momentum dispersion relation. These trends continue 
at 1.5 $\tc$, with a significant bound state showing up with 
a modified dispersion relation.

The $\jpsi$ moving in the heatbath frame sees more energetic gluons than 
the $\jpsi$ at rest. The fact that $\jpsi$ moving at such momenta show a
statistically significant peak indicates a stongly nonperturbative nature of
the plasma. We would like to clarify that we are investigating here $\jpsi$
states in an equilibriated plasma. This is very different from earlier
phenomenological studies concerned with survival of $\jpsi$ with large
trasnverse momentum from a finite sized plasma, where a large transverse
momentum may lead to the $\jpsi$ spending a smaller time in the plasma,
resulting in a larger survival probability \cite{chu,karsch}. However, such
studies also need to take into account the fact that the in-medium dispersion
relation for the $\jpsi$ will be modified. A modified dispersion relation
may also explain the large temperature dependence observed in the 
screening mass of the $\jpsi$ and the $\ec$ \cite{prd}; however, a
quantitative statement will require a much more exact determination of the
dispersion relation than is available at present.

%
%
%
%

\end{document}